\documentclass[%p
reprint,
 amsmath,amssymb,
 aps,
]{revtex4-2}

\usepackage{graphicx}% Include figure files
\usepackage{dcolumn}% Align table columns on decimal point
\usepackage{bm}% bold math
\usepackage{hyperref}% add hypertext capabilities
\usepackage{xcolor}
\newcommand{\bal}{\begin{align}}
\newcommand{\eal}{\end{align}}
\newcommand{\beq}{\begin{eqnarray}}
\newcommand{\eeq}{\end{eqnarray}}
\newcommand{\nneeq}{\nonumber \end{eqnarray}}

\newcommand{\es}{& = &}

\begin{document}

\preprint{APS/123-QED}

\title{Spiral flow of quantum quartic oscillator with energy cutoff}

\author{M. Girgu\'s}
 \email{mgirgus@fuw.edu.pl}
\author{S. D. G{\l}azek}%
 \email{stglazek@fuw.edu.pl}
 \affiliation{%
 Faculty of Physics, University of Warsaw, Warsaw, 02-093, Poland
}%

\date{April 26, 2024}% It is always \today, today,

\begin{abstract}
Theory of the quantum quartic oscillator is developed with close 
attention to the energy cutoff one needs to impose on the system 
in order to approximate the smallest eigenvalues and corresponding 
eigenstates of its Hamiltonian by diagonalizing matrices of limited
size. The matrices are obtained by evaluating matrix elements of 
the Hamiltonian between the associated harmonic-oscillator 
eigenstates and by correcting the computed matrices to compensate 
for their limited dimension, using the Wilsonian renormalization-group 
procedure. The cutoff dependence of the corrected matrices is found 
to be described by a spiral motion of a three-dimensional vector. 
This behavior is shown to result from a combination of a limit-cycle 
and a floating fixed-point behaviors, a distinct feature of the foundational 
quantum system that warrants further study. A brief discussion of the
research directions concerning renormalization of polynomial interactions
of degree higher than four, spontaneous symmetry breaking and coupling
of more than one oscillator through the near neighbor couplings known 
in condensed matter and quantum field theory, is included.
\end{abstract}

\maketitle

%%%%%%%%%%%
\section{Introduction}
%%%%%%%%%%%
The quantum quartic-oscillator Hamiltonian, 
\begin{equation}
\label{H}
H = -\frac{d^2}{d \varphi^2} + A \, \varphi^2 + B \, \varphi^4 \ ,
\end{equation}
is a foundational model for physics of many systems, ranging in 
scale from the particle~\cite{BW,[Status of Higgs Boson Physics in ]PDG} 
to atomic phenomena~\cite{CT} to condensed matter features~\cite{KGW} 
and cosmological theory~\cite{LR}. The coefficient $B$ of the quartic
term must be positive for the spectrum of $H$ to be bounded from 
below. Such positive quartic interaction term rapidly grows with the
magnitude of $\varphi$. In order to describe its effects in terms of 
the matrix elements of $H$ between the eigenstates of its harmonic 
part, $-d^2/d \varphi^2 + A \, \varphi^2$, one would need to consider
matrices of infinite size. The reason is that the energy that is quartic 
in $\varphi$ becomes infinitely greater than the quadratic energy when 
$\varphi$ grows to infinity. Instead, one can consider computations with 
the finite Hamiltonian matrices that are limited to the basis states whose 
harmonic energy does not exceed some ultraviolet cutoff. Then the question 
arises how the limited Hamiltonian matrix ought to be corrected, since $H$ 
in Eq.~(\ref{H}) corresponds to no such limitation. The answer is of interest 
in all areas of physics mentioned above and, by inference regarding the
method, wherever one works with an energy cutoff, or some similar cutoff, 
on the space of states.

The question was addressed by W\'ojcik~\cite{KWojcik} using the 
renormalization group (RG) procedure~\cite{KGW1965,KGW1970}. 
He numerically computed ${10 \times 10}$ matrices whose lowest 
eigenvalues accurately matched the lowest eigenvalues of the quartic 
oscillator matrices of size on the order of ${200 \times 200}$. He found 
that the computed matrices depend in a peculiar way on the number 
$n$ of rows and columns as it is lowered from 200 to 10. Namely, the 
evolving matrix elements rapidly move to a region of their nearly stable 
values and subsequently slowly drift away. This RG behavior requires 
understanding. We demonstrate below that it results from a limit-cycle 
behavior combined with an attraction to a floating fixed point. 

Since the Hamiltonian in Eq.~(\ref{H}) is used in testing, illustrating 
and explaining methods of solving quantum problems, as exemplified 
in~\cite{BW,PDG,CT,KGW,LR,KWojcik}, 
we should point out that the RG behavior gets quite intricate when 
the coefficient $A$ is allowed to be negative or when one adds to $H$ 
terms with higher powers of $\varphi$ than 4. However, our detailed 
discussion only concerns the case of Eq.~(\ref{H}) with $A > 0$.
We briefly comment on the more complex cases toward the end of
the paper.
%%%%%%%%%%%%%%
\section{Hamiltonian matrices}
%%%%%%%%%%%%%%
Following W\'ojcik~\cite{KWojcik}, we write the Hamiltonian of 
Eq.~(\ref{H}) in a dimensionless form
\begin{equation}
\label{Ha}
H = a^\dagger a + g(a^\dagger + a)^4 \ ,
\end{equation}
where $a$ and $a^\dagger$ denote the familiar annihilation 
and creation operators that satisfy the commutation relation 
${[a,a^\dagger]=1}$. The Hamiltonian $H$ of Eq.~(\ref{Ha}) 
provides energy in units of ${\hbar \omega}$. We omit the 
number 1/2 that shifts all eigenvalues equally. The normalized 
eigenstates ${|k\rangle = (k!)^{-1/2} a^{\dagger k} | 0 \rangle}$ 
of the term ${a^\dagger a}$, with eigenvalues $k$, are used 
to obtain the Hamiltonian matrix $H^\infty$ whose matrix 
elements are ${H^\infty_{k,l} = \langle k|H|l\rangle}$, where 
$k$ and $l$ range from 0 to $\infty$. 

In order to learn what form an effective 
Hamiltonian matrix with a finite cutoff $n$ on $k$ and 
$l$ should have, one starts with a cut off matrix 
$H^N$ of matrix elements  ${H_{k,l}^N = \theta(N-k) 
\theta(N-l) \, H_{k,l}^\infty}$, where $\theta$ is the 
Heaviside function and $N$ the cutoff. Subsequently, 
one eliminates, or ``integrates out'' a row and a column  
of the matrix eigenvalue problem for $H^N$ using the 
Gaussian elimination. Such elimination step produces 
a matrix with a cutoff $N-1$. It is an elementary form 
of the Wilsonian renormalization group transformation 
(RGT)~\cite{KGW1970}. The goal is to repeat the RGT 
many times and obtain matrices with cutoffs $N-2$, 
$N-3$ and so on until one reaches $n$. In the process 
one learns how the matrix with the small cutoff $n$  
is related to the initial matrix with a large cutoff $N$.

After $N-n$ rows and columns are so eliminated, the 
resulting matrix with cutoff ${n \ll N}$ is denoted by 
$H_n^N$. The cutoff $n$ is called the floating 
cutoff~\cite{[See Sec. 12.4 in ]Weinberg}, even 
though it changes in discrete steps. The name is adequate 
for ${n \gg 1}$ because the cutoff change in every step is
small in comparison with the cutoff itself and the RGT
appears to the eye as nearly continuous. 

By construction, 
the eigenvalues ${E \ll n}$ of the matrix $H_n^N$ do not 
depend on the floating cutoff $n$. Thus, in the limit 
${N \to \infty}$ one obtains the renormalized Hamiltonian 
matrices \cite{[See Fig. 6 in Sec. VII B of ]Wilsonetal}
\begin{equation}
H_R^n = \lim_{N \to \infty} H_n^N \ , 
\end{equation}
whose eigenvalues ${E \ll n}$ also do not depend on the 
floating cutoff $n$. In the quartic-oscillator case 
there is no need to counter divergences, which simplifies
the RG procedure in comparison with models that involve
divergences and require computation of the corresponding 
counter terms.

Numerical results available in Ref.~\cite{KWojcik} are 
for ${N \sim 200}$ and $n$ between $10$ and $N$. For 
example, matrix $H_R^{10}$ with ${g \sim 10}$, 
approximated numerically by matrices $H_{10}^{200}$, 
reproduce the lowest eigenvalue of $H^{200}$ with 
accuracy ${\sim 0.35\%}$. In contrast, plainly cut off 
matrix $H^{10}$ produces a 100 times greater error. 

The benefit of computing the renormalized Hamiltonian 
matrix of size ${n \times n}$ is that one can use it instead 
of matrices ${N\times N}$ for approximate description of 
the quartic-oscillator in interaction with some other system.
The method can work provided that the external interaction 
does not significantly excite the oscillator states that lie 
outside the range of the floating cutoff $n$.

Since the quartic interaction term only changes the 
number of quanta by 0, 2 or 4, the real and symmetric 
Hamiltonian matrix $H^N$ has non-zero matrix elements 
only in a band formed by the diagonal and four closest 
non-vanishing near-diagonals. In consequence, the only 
matrix elements of $H_n^N$ that differ from the matrix 
elements of the initial matrix $H^N$ with the same 
subscripts, lie in the corner of the former with subscripts 
$k$ and $l$ equal $n$ or ${n-2}$. In addition, the Gaussian 
elimination integrates out even rows and columns in the 
eigenvalue equation of $H_n^N$ similarly to but independently 
of how it integrates out the odd ones. The result is that the 
variation of $H_n^N$ with $n$ can be parameterized using 
only three numbers ${\xi_i, i = 1,2,3}$. Namely,
\begin{subequations}
\label{xidef}
\begin{align}
\label{xidef1}
&H_{n ;  n,n}^N = n + \xi_1 \left( H^N_{n,n} - n \right) \ , 
\\
\label{xidef2}
&H_{n ;  n-2,n-2}^N = n-2  + \xi_2 \left[ H^N_{n-2,n-2} - (n-2) \right] \ ,
\\
\label{xidef3}
&H_{n ;  n,n-2}^N = \xi_3 H^N_{n,n-2} \ ,
\\
\label{xidef4}
&H_{n ;  n-2,n}^N = \xi_3 H^N_{n-2,n} \ ,
\end{align}
\end{subequations}
where the semicolon separates the floating cutoff $n$ 
from the matrix element subscripts. The numbers
$\xi_1$, $\xi_2$ and $\xi_3$ are the ratios of the 
interaction matrix elements evolving with cutoff $n$ to the 
original ones with the same subscripts. The cutoff-flow 
of the Hamiltonian matrices that correspond to Eq.~(\ref{H})
is thus fully described by a sequence of three-dimensional 
vectors $\vec \xi(n)$, ${n = N, N-2, N-4}$ and so on.

Reference~\cite{KWojcik} identifies a universal feature of the 
sequences $\vec \xi(n)$ for various choices of the coupling 
constant $g$ and the initial vector $\vec \xi(N)$. Sequences 
$\vec \xi(n)$ rapidly approach the vicinity of about ${(0.2,0.8,
0.5)}$ and subsequently all three components of $\vec \xi$ 
appear to vary quite slowly. Nevertheless, they steadily increase 
while $n$ decreases down to the values on the order of the 
eigenvalue for which the sequence is generated. The nature 
of this behavior is not explained in~\cite{KWojcik}. We report 
the finding that the numerically observed sequences correspond
to a spiral RG behavior that results from an interplay between 
the limit-cycle~\cite{KGWlimitcycle,SDGKGW} and floating 
fixed-point behaviors.
%%%%%%%%%%%%%%%%%%%%%%%
\section{ Fixed points and limit cycle }
%%%%%%%%%%%%%%%%%%%%%%%
Straightforward algebra shows that the RGT, or the recursion 
one obtains by applying the Gaussian elimination to the 
eigenvalue problem for the matrix $H^N$, is described 
by the following equations,
\begin{subequations}
\label{rt}
\begin{align}
\label{rt1}
&\xi_1(n-2) = \xi_2(n) - \phi_1(n) \, \xi_3(n)^2/d(n) \ ,
\\
&\label{rt2}
\xi_2(n-2) = 1 - \phi_2(n) /d(n)  \ ,
\\
&\label{rt3}
\xi_3(n-2) = 1 - \phi_3(n) \, \xi_3(n) /d(n) \ , 
\end{align}
\end{subequations}
with the denominator
\beq
\label{d}
d(n) \es \xi_1(n)+ (n-E)/[g \phi(n)] \ ,
\eeq
and
\begin{subequations}
\begin{align}
&\phi_1(n) = \frac{n(n-1)(4n-2)^2}{(6n^2-18n+15)\phi(n)} \ ,
\\
&\phi_2(n) = \frac{n(n-1)(n-2)(n-3)}{(6n^2-42n+75)\phi(n)} \ ,
\\
&\phi_3(n) = \frac{n(n-1)(4n-2)}{(4n-10)\phi(n)} \ ,
\\
&\phi(n) = 6n^2+6n+3 \ .
\end{align}
\end{subequations}
%%%%%%%%%%%%%%%%%%%
\subsection{ Fixed points and their confluence }
\label{fp+}
%%%%%%%%%%%%%%%%%%%
When the eigenvalue $E$ is very small in comparison 
to a  large  ${n \sim N \to \infty}$, the equations 
that describe the RGT  take the form
\begin{subequations}
\label{largen}
\begin{align}
\label{largen1}
&\xi_1(n-2) = \xi_2(n) - 4 \ \xi_3(n)^2 / [9 \, d_s(n)] \ ,
\\
\label{largen2}
&\xi_2(n-2) = 1 - 1/[36 \, d_s(n)] \ ,
\\
\label{largen3}
&\xi_3(n-2) = 1 - \xi_3(n) / [6 \, d_s(n)]  \ ,
\end{align}
\end{subequations}
with the simplified denominator, 
\beq
\label{ds}
d_s(n) \es \xi_1(n) + 1/(6gN) \ ,
\eeq
and ${N-n}$ neglected in comparison to $N$ and $n$. 
The number $1/(6gN)$ in $d_s(n)$ is retained because 
$\xi_1$ can {\it in principle} be arbitrarily small. Writing 
Eqs.~(\ref{largen}) in the form,
\begin{equation}
\label{vecF}
\vec \xi(n-2) = \vec F[\vec \xi(n)] \ ,
\end{equation}
one introduces the rational vector function $\vec F$ of $\vec \xi$.  
Fixed points of the transformation in Eq.~(\ref{vecF}), denoted
by $\vec \xi^*$, are defined as solutions to the equation
\begin{equation}
\vec F[ \vec \xi^*] = \vec \xi^* \ .
\end{equation}
There are two such solutions, $\vec \xi^* = \vec \xi^\pm$, where
\begin{subequations}
\label{xi+-}
\begin{align}
\label{xi+}
&\vec \xi^+ = \left( -\frac{1}{6gN} + \frac{1}{6} \frac{1+p}{1-p}, 1 - \frac{1}{6} \frac{1-p}{1+p}, \frac{1}{2} + \frac{p}{2} \right) \ ,
\\
\label{xi-}
&\vec \xi^- = \left( -\frac{1}{6gN} + \frac{1}{6} \frac{1-p}{1+p}, 1 - \frac{1}{6} \frac{1+p}{1-p}, \frac{1}{2} - \frac{p}{2} \right) \ ,
\end{align}
\end{subequations}
with ${p^2 = \sqrt{ a + (a/2)^2} - a/2}$ and ${1/a = 4gN}$. 
The fixed point $\vec \xi^+$ turns out to be attractive
while $\vec \xi^-$ is repulsive. Sending $N$ to infinity, so that 
$p$ tends to zero, leads to confluence of the two fixed points 
into one,
\begin{equation}
\vec \xi^+ = \vec \xi^-  = (1/6, 5/6, 1/2) \ . 
\end{equation}
The three components of this vector qualitatively explain 
the magnitudes of numbers ${(0.2,0.8,0.5)}$ and nearly 
fixed-point behavior of $\vec \xi(n)$ found numerically 
in Ref.~\cite{KWojcik}. However, for non-zero values 
of $p$, the two fixed points are separate. Next section 
describes the behavior of $\vec \xi$ in the vicinity of 
$\vec \xi^+$.
%%%%%%%%%%%%%%%%%%%%%%%%%%%%%%
\subsection{ RG evolution near fixed point $\vec \xi^+$}
\label{nearfixed}
%%%%%%%%%%%%%%%%%%%%%%%%%%%%%%
Using formula ${\vec \xi(n) = \vec \xi^+ + \Delta \vec \xi(n)}$
and keeping only terms linear in ${\Delta \vec \xi}$ one obtains 
the recursion
\begin{equation}
\label{actionF'}
\Delta \vec \xi (N-2k) = [\vec F'[\vec \xi^+] ]^{k}\ \Delta \vec \xi(N) \ ,
\end{equation}
where 
\beq
\label{Fprim}
\vec F' {[} \vec \xi^+ {]} \es r \, G
\eeq 
is the derivative of $\vec F$, the ratio $r =(1-p)/(1+p)$ 
and the matrix $G$ reads
\begin{eqnarray}
G 
& = &
\left[ \begin{array}{ccc}
4(1-p^2) & \frac{1+p}{1-p} & -\frac{8}{3}(1+p) \\
\frac{1-p}{1+p} & 0 & 0 \\
3 (1-p) & 0 & -1
\end{array} \right] \ .
\end{eqnarray}
The eigenvalues of $G$ are ${\lambda = 1}$ and ${\lambda_\pm
= e^{\pm i\omega}}$ with ${\omega = 2 \arcsin p}$.
The corresponding eigenvectors are
\begin{subequations}
\label{eigenvalues1}
\begin{align}
&\lambda \ \rightarrow \ \vec v =
\left[ \frac{2}{3} \frac{1}{1- p}, \frac{2}{3} \frac{1}{1 + p}, 1 \right] , 
\\
&\lambda_\pm \ \rightarrow \ \vec v_\pm = \vec v_1 \pm i \vec v_2 =
\left[  \frac{2}{3} \frac{e^{\pm i \omega/2}}{1-p} , 
         \frac{2}{3} \frac{e^{\mp i \omega/2}}{1+p} , 
        \frac{1}{\sqrt{1-p^2}} \right] .
\end{align}
\end{subequations}
The real vectors $\vec v$,
\begin{subequations}
\begin{align}
&\vec v_1 = \left( \frac{2}{3} \frac{\sqrt{1-p^2}}{1-p}, \frac{2}{3} \frac{\sqrt{1-p^2}}{1+p}, \frac{1}{\sqrt{1-p^2}} \right) \ ,
\\
&\vec v_2 = \left( \frac{2}{3} \frac{p}{1-p}, - \frac{2}{3} \frac{p}{1+p}, 0 \right) \ ,
\end{align}
\end{subequations}
are linearly independent.
An arbitrary vector $\Delta \vec \xi$ can be represented as
\begin{equation}
\label{basis}
\Delta \vec \xi = \alpha \vec v + \beta \vec v_1 + \gamma \vec v_2 \ ,
\end{equation}
where the coefficients are
\begin{subequations}
\label{alphabetagamma}
\begin{align}
\label{alpha}
&\alpha = \frac{3}{4p^2} \left[ (1-p) \Delta \xi_1 + (1+ p) \Delta \xi_2 \right] - \frac{1-p^2}{ p^2} \Delta \xi_3 ,
\\
\label{beta}
&\beta = \frac{\sqrt{1-p^2}}{p^2} \left\{ \Delta \xi_3 - \frac{3}{4} \left[ (1-p) \Delta \xi_1 + (1+p) \Delta \xi_2 \right] \right\} ,
\\
\label{gamma}                   
&\gamma  = \frac{3}{4p} \left[ (1-p) \Delta \xi_1  -  (1+ p) \Delta \xi_2 \right] .
\end{align}
\end{subequations}
These coefficients diverge for ${p \to 0}$. However, their diverging
values are countered by the smallness of $\Delta \vec \xi$ when 
one considers RGT in the vicinity of $\vec \xi^+$.

Action of the matrix $G$ on the vector $\Delta \vec \xi$,
represented in terms of the coordinates ${(\alpha,\beta,\gamma)}$ 
in the basis of ${\{ \vec v, \vec v_1, \vec v_2\}}$, is described by
the formula
\begin{eqnarray}
\label{alphabetagammaevolution}
G \ \left[ \begin{array}{c} \alpha \\ \beta \\ \gamma \end{array} \right]  & = & 
\left[ \begin{array}{ccc}
1  & 0 & 0 \\
0 & \cos \omega & \sin \omega \\
0 &-\sin \omega & \cos \omega
\end{array} \right] \left[ \begin{array}{c} \alpha \\ \beta \\ \gamma \end{array} \right] \ .
\end{eqnarray}
Repeated action of $G$ thus yields a cyclic behavior of coordinates 
$\beta$ and $\gamma$ as functions of the floating cutoff $n$ with
period $t = {2 \pi / \omega}$. 

Since ${p \sim (4 g N)^{-1/4}}$, see the formula below 
Eq.~(\ref{xi-}), the period of cyclic evolution of $(\beta,\gamma)$, 
which is ${t \sim \pi/p \sim \pi (4 g N)^{1/4}}$, is much smaller 
than $N$ for large $N$. This means that the number of $(\beta,\gamma)$ 
cycles produced near $\vec \xi^+$ by the RGT of Eq.~(\ref{actionF'}), 
repeated $k=(N-n)/2$ times, is large and increases with $N$ while 
$n$ is fixed.

Besides the matrix $G$, the RGT derivative $\vec F'[\vec \xi^+]$ 
of Eq.~(\ref{Fprim}) includes the factor ${r = (1-p)/(1+p) < 1}$, 
which changes the cycle to a spiral. We call $r$ the spiral 
convergence factor. Its presence implies that $|\Delta \vec \xi\,|$ 
contracts at the rate of $\sim r^t$ per cycle. In the scaling 
coordinates $r^{-k} \beta$ and $r^{-k} \gamma$, the 
approximate RG spiral corresponds to a circle.

We explain details of the evolution of $\vec \xi$ 
and illustrate their characteristic features by some figures 
in the sections that follow. Our explanation includes the 
small drift of $\vec \xi$ with $n$, which was previously 
observed numerically in Ref.~\cite{KWojcik} and required 
understanding. The drift results from a small variation of 
the formula for $\vec F[\vec \xi]$ as the floating cutoff 
$n$ changes. 
%%%%%%%%%%%%%%%%%%%%%%%
\subsection{Floating fixed-point $\vec \xi^+(n)$}
\label{ffp+}
%%%%%%%%%%%%%%%%%%%%%%%
The exact RGT of Eqs.~(\ref{rt}) and its simplified 
form given in Eqs.~(\ref{largen}) for large floating 
cutoffs $n$ are slightly different. Namely, the exact 
RGT depends on the floating cutoff $n$ itself. Therefore,
instead of a fixed point, it actually produces sequences 
$\vec \xi(n)$ that converge to the sequence we denote 
by $\vec \xi^+(n)$ and for brevity call the attractive 
sequence. 

Due to the weak dependence of the RGT on $n$, one 
can estimate $\vec \xi^+(n)$ using an approximation 
${\vec \xi_a^+(n-2) \approx \vec \xi_a^+(n)}$ in 
Eqs.~(\ref{rt}). The resulting set of equations for 
the sequence $\vec \xi_{a}^+(n)$, 
\begin{subequations}
\begin{align}
\label{xi1plusapprox}
&\xi_{a,1}^+(n) = \xi_{a,2}^+(n) - \phi_1(n) \, \xi_{a,3}^+(n)^2 / d_a(n) \ ,
\\
\label{xi2plusapprox}
&\xi_{a,2}^+(n) = 1 - \phi_2(n) / d_a(n)  \ ,
\\
\label{xi3plusapprox}
&\xi_{a,3}^+(n) = 1 - \phi_3(n) \, \xi_{a,3}^+(n) / d_a(n)  \ , 
\end{align}
\end{subequations}
with the approximated denominator, 
\beq
\label{da}
d_a(n) \es \xi_{a,1}^+(n) + (n-E)/[g \phi(n)] \ ,
\eeq
typically has two real vector solutions. That number can 
only shrink to one or zero for small values of $n$. The 
fixed point solutions of  Eqs.~(\ref{xi+-}) suggest that 
the vector solution $\vec \xi_{a}^+(n)$ that corresponds
to the attractive sequence $\vec \xi^+(n)$ for $n \gg E$ 
is the one of the two solutions that has larger components.

Independent numerical computation shows that $\vec \xi_{a}^+(n)$ 
approximates the actual attractive floating fixed-point solution 
$\vec \xi^+(n)$ with relative error smaller than ${5 \%}$ for ${n \geq 6}$ 
when ${g=1}$ and ${E=0}$. The accuracy of approximation 
increases with increasing $n$ and the relative error becomes 
smaller than ${1 \%}$ for ${n \geq 80}$. The values of ${g}$ 
and ${E}$ used here are suitable for discussion of generic 
features of the sequences $\vec \xi^+(n)$ generated numerically. 

Sequences $\vec \xi(n)$ 
generated by the exact RGT tend to $\vec \xi^+(n)$ while $n$ 
is large. Difference between $\vec \xi(n)$ and $\vec \xi^+(n)$ 
decreases with increasing ${N-n}$, because the converging 
spiral evolution develops over more periods. Exceptions to such 
convergence result from existence of another special sequence, 
called repulsive, which we discuss in Sec.~\ref{ffp-}.

The attractive sequence $\vec \xi^+(n)$ can be described 
in words as a floating center of convergence for the exact
spiral $\vec \xi(n)$. The floating spiral-center 
explains the RG evolution of the quartic oscillator observed 
in~\cite{KWojcik}. The rapid approach of $\vec \xi$ to the 
vicinity of ${(0.2, 0.8, 0.5)}$ corresponds to the convergence 
of $\vec \xi(n)$ to $\vec \xi^+(n)$ as $n$ decreases. But 
after some initial RGT steps, $\vec \xi(n)$ spirals around 
$\vec \xi^+(n)$ so closely that the spiral is not visible to 
the naked eye in the figures provided in~\cite{KWojcik}. 
Instead, the sequences $\vec \xi(n)$ appear in~\cite{KWojcik} 
to slowly and steadily increase in some direction as $n$ 
decreases continuing to obey the condition $n \gg E$.
When the magnitude of $n$ approaches that of $E$, rapid 
changes of $\vec \xi$ occur whose variety depends on $g$ 
and $E$ one considers. The rapid changes of $\vec \xi$ for 
${n \leq E}$ are not visible in the figures of~\cite{KWojcik}, 
because the plots there are generated for ${E=0}$.

\begin{figure}[ht!]
\includegraphics[scale=0.20]{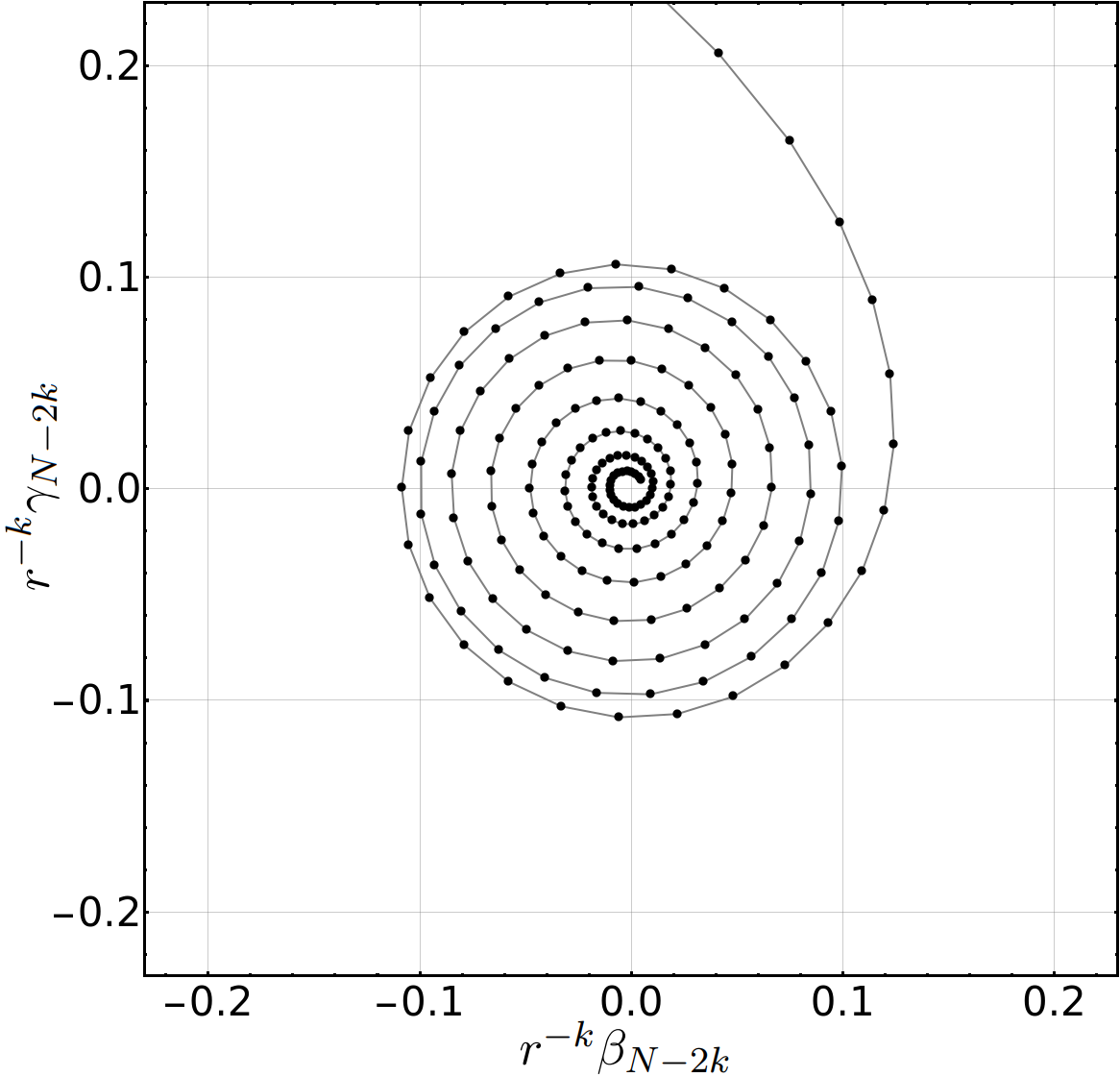} 
\caption{ RG evolution of vector $\vec \xi(n)$ near the 
attractive nearly-fixed point $\vec \xi^+(n)$ for the 
running energy cutoff $n = N-2k$ and $k$ ranging 
from $8$ to $200$. Coordinates $\beta$ and $\gamma$ 
are defined in Eqs.~(\ref{beta}) and (\ref{gamma}) 
for ${\Delta \vec \xi (N-2k) = \vec \xi (N-2k) - \vec \xi^+(N-2k)}$. 
The initial cutoff is ${N=1000}$ and ${\vec \xi_{N}
=\left(1,1,1\right)}$. The floating fixed point 
$\vec \xi^+(N-2k)$ is derived numerically by applying 
the exact RGT of Eqs.~(\ref{rt}) to ${\vec \xi_{N^\prime}
=\left(1,1,1\right)}$ with ${N^\prime = 1200}$, while 
${E=0}$ and ${g=1}$. The points do not lie on a circle, 
as they would in case of the approximate RGT of 
Eqs.~(\ref{largen}),  because the rate of spiral 
convergence per cycle in the exact RGT is 
slightly greater than $r^t$. The consecutive points 
are connected to guide the eye.}
\label{fig1}
\end{figure}

One can study behavior of sequences $\vec \xi(n)$ near 
$\vec \xi^+(n)$ using formula ${\vec \xi (n) = \vec \xi^+(n) 
+ \Delta \vec \xi (n)}$ and expanding the recursion 
of Eqs.~(\ref{rt}) in a series of powers of $\Delta \vec \xi (n)$,
in analogy to the procedure used in Sec.~\ref{nearfixed}.
The linear terms, dominant for small $\Delta \vec \xi (n)$, 
obey the recursion
\begin{equation}
\label{actionF'n}
\Delta \vec \xi (n-2) = \vec F'_n [\vec \xi^+(n)] \Delta \vec \xi (n) \ .
\end{equation}
The matrix $\vec F'_n [\vec \xi^+(n) ]$ explicitly depends 
on $n$. Its eigenvalues and eigenvectors are not constant 
during the RG evolution. Nevertheless, their change with 
$n$ is slow for large $n$. The numerically obtained spiral convergence of sequences 
$\vec \xi(n)$ to $\vec \xi^+(n)$ is illustrated in Fig.~\ref{fig1}, 
where we plot points with coordinates 
\beq
\label{bg}
(\beta,\gamma)_{\rm scaling} \es {\left( r^{-k} \beta_{N-2k}, 
r^{-k} \gamma_{N-2k} \right)}
\eeq
in the same basis as in Eq.~(\ref{basis}). The factor $r^{-k}$ cancels 
the factor $r^k$ that is present in Eq.~(\ref{actionF'}). 

In case of the approximate RGT of Eqs.~(\ref{largen}), the 
points plotted in Fig.~\ref{fig1} would lie on a circle, as predicted 
by Eq.~(\ref{alphabetagammaevolution}) and commented on
in the one before the last paragraph of Sec.~\ref{nearfixed}. 
In case of the exact RGT of Eqs.~(\ref{rt}), the sequence ${r^{-k} 
\Delta \vec \xi (n=N-2k)}$ slowly spirals to 0 as $k$ increases, 
because the exact RGT spiral convergence factor analogous to $r$ 
in Eq.~(\ref{Fprim}) differs from the constant $r$; it slowly decreases 
when $n$ decreases. In addition, the distance between points 
${\left( r^{-k} \beta_{N-2k}, r^{-k} \gamma_{N-2k} \right)}$ and 
the center of Fig.~\ref{fig1} doesn't decrease monotonically. This 
feature is caused by a drift of eigenvectors $\vec v$, $\vec v_1$, 
$\vec v_2$ with decreasing $n$, which is discussed in 
Sec.~\ref{MonotonicityOfSpirals}. 
%%%%%%%%%%%%%%%%%%%%%%%%%%%%%
\subsection{Variation of the RG spiral with floating cutoff $n$}
\label{MonotonicityOfSpirals}
\begin{figure}[ht!]
\includegraphics[scale=0.242]{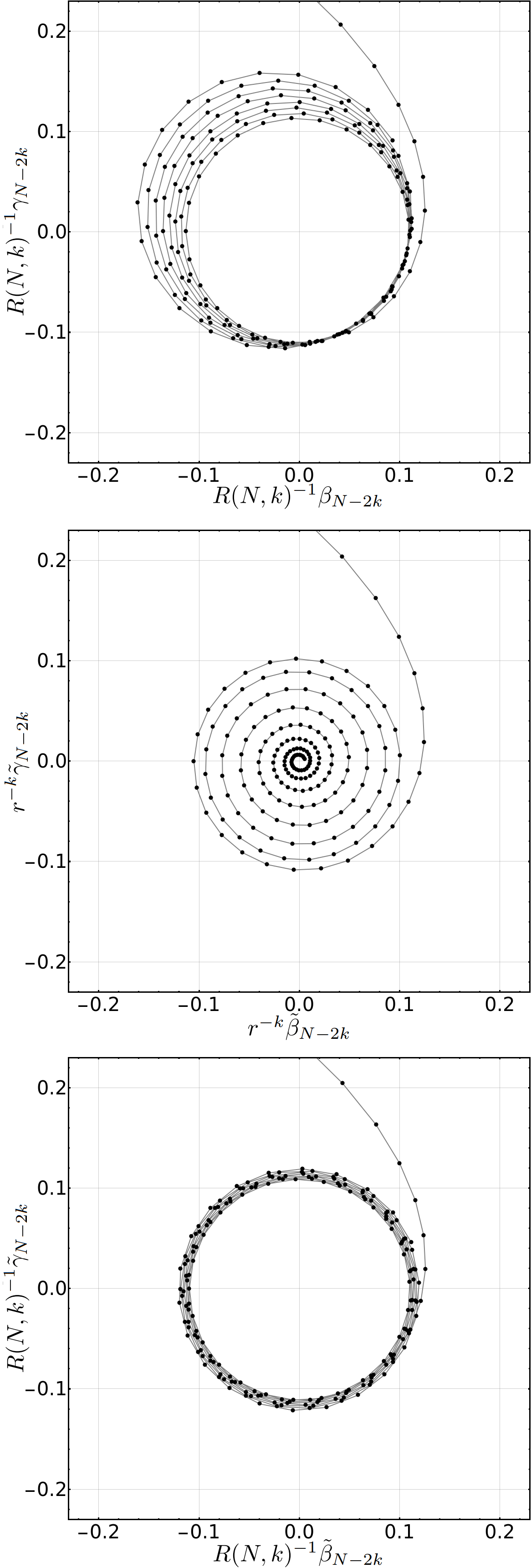} 
\caption{
The same RG spiral as in Fig.~\ref{fig1} displayed in terms 
of three different coordinate systems, from top to bottom: 
${( \beta, \gamma )_{\rm scaling \ 1}}$ of Eq.~(\ref{topcoor}),
${( \beta, \gamma )_{\rm scaling \ 2}}$ of Eq.~(\ref{middlecoor})
and 
${( \beta, \gamma )_{\rm scaling \ 3}}$ of Eq.~(\ref{bottomcoor}), 
correspondingly. Coordinates used in the top figure account for 
the cutoff flow of the spiral convergence rate $r$. The middle 
figure shows the effect of variation of the basis vectors formed 
by the eigenvectors of the RGT derivative near the floating 
fixed-point $\vec \xi^+(n)$. The bottom figure shows the result 
of simultaneous account for the both effects. }
\label{fig2}
\end{figure}

%%%%%%%%%%%%%%%%%%%%%%%%%%%%%
It is pointed out at the end of previous subsection that the matrix 
$\vec F'_n [\vec \xi^+(n)]$ in Eq.~(\ref{actionF'n}) depends on the 
floating cutoff $n$ so that the exact RGT produces a non-monotonous 
spiral flow in Fig.~\ref{fig1}, instead of the circle that the simplified 
Eq.~(\ref{alphabetagammaevolution}) would yield in terms of the
coordinates $(\beta,\gamma)_{\rm scaling}$. This feature is a consequence of the 
fact that the moduli of eigenvalues of $\vec F'_n [\vec \xi^+(n)]$ don't 
have a constant value equal to $r$. The rate of spiral convergence per 
one RGT step is instead given by some $n$-dependent value $r_n$. 
For large $n$, $r_n$ is given by the same formulas as $r$
except that the fixed $N$ is replaced by the floating $n$,
\begin{equation}
r_n = \frac{1-p_n}{1+p_n} \ ,
\end{equation}
where ${p_n^2 = \sqrt{ a_n + (a_n/2)^2} - a_n/2}$ and ${1/a_n = 4gn}$.
The convergence factor after $k$ RGT steps is thus given by
\begin{equation}
\label{RNk}
R(N,k) = r_{N-2} \cdot \dotsc \cdot r_{N-2k} \ ,
\end{equation}
which replaces $r^k$.
The difference between $r^k$ and varying convergence factors 
$R(N,k)$ is illustrated in the top panel of Fig.~\ref{fig2}, where 
we plot points with coordinates 
\beq
\label{topcoor}
(\beta,\gamma)_{\rm scaling \ 1} \es 
R(N,k)^{-1}  \left( \beta_{N-2k}, \gamma_{N-2k} \right) \ . 
\eeq
Distance of these points from ${\left( 0, 0 \right)}$ is roughly constant 
for $\beta>0$ and $\gamma<0$, and it increases with increasing number 
of cycles for points lying in the first, second and third quadrants of the 
coordinate system.

Variation with $n$ of the eigenvectors $\vec v$, $\vec v_1$ 
and $\vec v_2$ of the derivative $\vec F'_n [\vec \xi^+(n)]$ is 
illustrated in the middle panel of Fig.~\ref{fig2}. One changes the 
coordinates $\alpha$, $\beta$, $\gamma$ correspondingly to the 
replacement of $p$ by $p_n$ in Eqs.~(\ref{alphabetagamma}). 
The new coordinates are denoted by $\tilde{\alpha}_n$, $\tilde{\beta}_n$, 
$\tilde{\gamma}_n$. The middle panel of Fig.~\ref{fig2} shows 
the RG flow of 
\beq
\label{middlecoor}
(\beta,\gamma)_{\rm scaling \ 2}
\es
r^{-k} \left( \tilde \beta_{N-2k}, \tilde \gamma_{N-2k} \right) \ .
\eeq
In this representation the spiral converges 
quite uniformly, as opposed to the somewhat 
erratic convergence illustrated in Fig.~\ref{fig1}, 
where one uses the fixed basis of 
Eqs.~(\ref{alphabetagamma}).

The bottom panel in Fig.~\ref{fig3} displays the
flow of $\Delta\vec \xi$ using coordinates
\beq
\label{bottomcoor}
(\beta,\gamma)_{\rm scaling \ 3}
\es
R(N,k)^{-1} \left( \tilde \beta_{N-2k}, \tilde \gamma_{N-2k} \right) \ .
\eeq
Somewhat erratic spiral flow in Fig.~\ref{fig1}, 
the angular asymmetry visible in the top panel 
of Fig.~\ref{fig2}, the regular spiral in the middle 
panel of Fig.~\ref{fig2} and the circle shown in 
the bottom panel of Fig.~\ref{fig2} together  
identify and summarize the main features of the 
RG spiral flow of the quartic oscillator Hamiltonian 
with the cutoff near the attractive sequence 
$\vec \xi^+(n)$. When the dependence of 
eigenvalues and eigenvectors of the derivative 
$\vec F'_n [\vec \xi^+(n)]$ is factored out, the 
scaling $\Delta \vec \xi(n)$ moves around a 
circle. The remaining apparent variation of the 
circle radius is not further discussed in this paper.
%%%%%%%%%%%%%%%%%%%%%%%
\subsection{Repulsive floating fixed-point $\vec \xi^-(n)$}
\label{ffp-}
%%%%%%%%%%%%%%%%%%%%%%%
The approximate RGT has a second fixed point solution $\vec \xi^-$ defined in 
Eq.~(\ref{xi-}). The corresponding eigenvalues of the derivative matrix $\vec F'[\vec \xi^-]$ are
$r^{-1}$, $r^{-1} e^{i \omega}$, $r^{-1} e^{- i \omega}$, with the 
same $r$ and $\omega$ as for $\vec \xi^+$. This implies that $\vec \xi^-$ 
is a repulsive fixed point. The sequences of $\vec \xi$ that start in the vicinity 
of $\vec \xi^-$, move away from $\vec \xi^-$ following a spiral curve. The rate 
of divergence per step is $r^{-1}$, which is an inverse of the rate of convergence 
on $\vec \xi^+$. The frequency with which the spiral unfolds around $\vec \xi^-$ 
is $\omega$, the same as in the case of folding in around $\vec \xi^+$.

Similarly to the attractive fixed point, the repulsive fixed point $\vec \xi^-$ 
of the approximate RGT has a counterpart in the exact RGT, which is a 
repulsive floating fixed-point $\vec \xi^-(n)$. The sequences repelled from 
$\vec \xi^-(n)$ can be equivalently described as the convergent sequences 
that are generated by the inverse RGT,
\begin{subequations}
\begin{align}
&\xi_1(n) = - \frac{n-E}{g \phi(n)} + \frac{\phi_2(n)}{1 - \xi_2(n-2)} ,
\\
&\xi_2(n) = \xi_1(n-2) + \frac{\phi_1(n) \phi_2(n)}{\phi_3^2(n)} \frac{\left[ 1 - \xi_3(n-2) \right]^2}{1 - \xi_2(n-2)} ,
\\
&\xi_3(n) = \frac{\phi_2(n)}{\phi_3(n)} \frac{1 - \xi_3(n-2)}{1 - \xi_2(n-2)} .
\end{align}
\end{subequations}
These sequences converge to $\vec \xi^-(n)$. 
The inverse RGT can be used to find numerical 
approximations for the repulsive floating fixed-point 
sequence.

\begin{figure}[h!]
\includegraphics[scale=0.28]{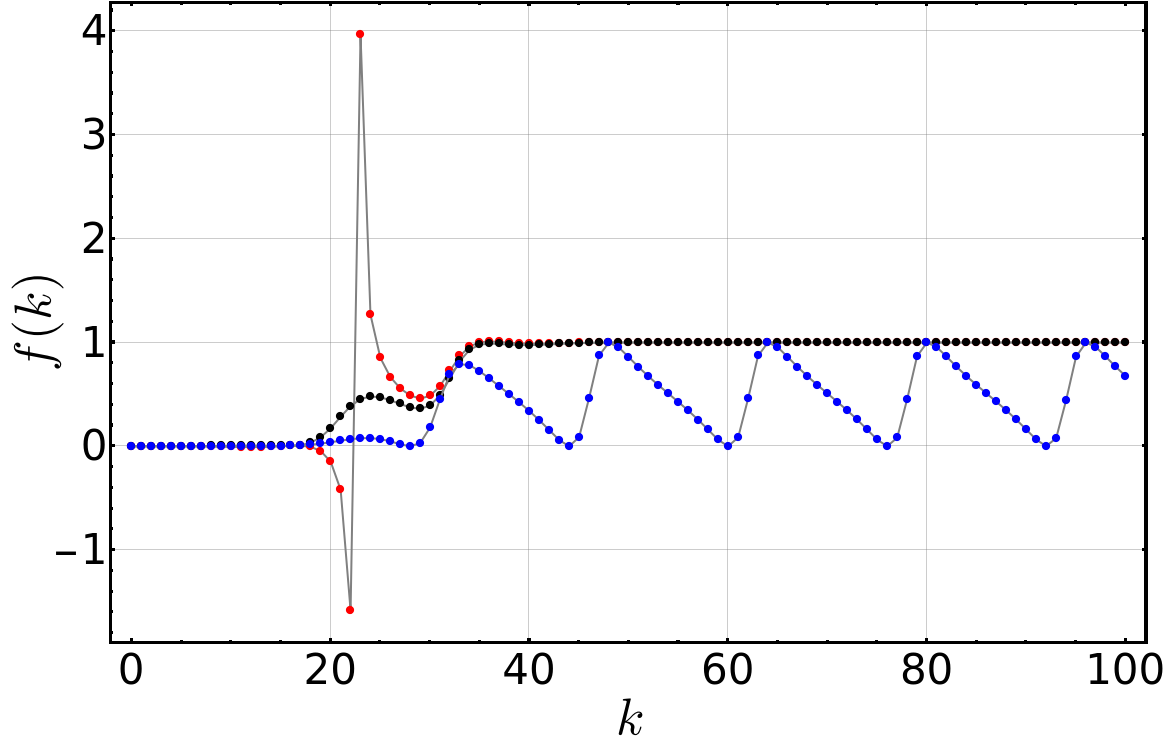} 
\caption{Examples of sequences of projections $f(k)$ of Eq.~(\ref{fk}), where $k$ is 
the number of RGT steps. Value ${f(k)=0}$ corresponds to projection of $\vec \xi(N-2k)$
being the same as for $\vec \xi^-$ and ${f(k)=1}$ means that the projection of 
$\vec \xi(N-2k)$ is the same as for $\vec \xi^+$. The points are generated by the 
approximate RGT with $6gN = 10^3$ and initial conditions 
$\vec \xi (N) = \vec \xi^- + (10^{-6}, 10^{-6}, 10^{-6})$ (black points), 
$\vec \xi (N) = \vec \xi^- + (10^{-6}, 10^{-6}, 2 \times 10^{-6})$ (red points) 
and $\vec \xi (N) = \vec \xi^-  + (10^{-6}, 10^{-6}, 1.5 \times 10^{-6})$ (blue points).
The points are connected to guide the eye.} 
\label{fig3}
\end{figure}

One might expect that the sequence $\vec \xi(n)$ 
that is repelled from $\vec \xi^-(n)$ goes over to the 
attractive sequence $\vec \xi^+(n)$. However, the 
sequences generated by the RGT from the initial conditions 
$\vec \xi (N)$ in the vicinity of the repulsive fixed point, 
do not always simply transit over to the attractive fixed 
point as it is the case for the initial condition ${\vec \xi (N) 
= (1,1,1)}$. We illustrate this finding in Fig.~\ref{fig3}, 
where plots of the projection
\begin{equation}
\label{fk}
f(k) = \left[ \vec \xi (N-2k) - \vec \xi^- \right] \cdot \frac{ \vec \xi^+ - \vec \xi^- }{\left( \vec \xi^+ - \vec \xi^- \right)^2} \ ,
\end{equation}
as a function of the number of RGT steps $k$, 
are shown for three different initial points $\vec \xi (N)$. 

The first sequence, represented by the black dots, approaches 
1, which corresponds to ${\vec \xi (N-2k) = \vec \xi^+}$, 
after just two turns. The second sequence, represented by the 
red points, makes a relatively big jump before reaching 1, 
which is caused by the approximate RGT denominator 
$d_s(n)={\xi_1(n) + 1/(6gN)}$ in Eq.~(\ref{ds}) becoming 
negative for some value of $n$ just before the jump. The 
third sequence, represented by the blue points, doesn't converge 
to 1 at all. 

The effects displayed in Fig.~\ref{fig3} are
due to a special property of the simplified RGT of Eqs.~(\ref{largen}). 
Namely, any two sequences generated by the approximate RGT, 
say $\vec \xi_A (n)$ and $\vec \xi_B (n)$, satisfy the equation
\begin{align}
\label{cone}
&h\left[ \vec \xi_A (n-2) - \vec \xi_B (n-2) \right] 
= 
\nonumber\\ &= \frac{h\left[ \vec \xi_A (n) - \vec \xi_B (n) \right]}
{36\left[ \xi_{A 1}(n) + 1/(6gN) \right] \left[ \xi_{B 1}(n) + 1/(6gN) \right]} \ ,
\end{align}
where ${h\left[ \vec v \right] = v_1 v_2 - (4/9) v_3^2}$.
Equation ${h\left[ \vec \xi - \vec \xi^- \right]=0}$ describes 
a cone in the three-dimensional space of $\vec \xi$. The
tip of the cone is in ${\vec \xi = \vec \xi^-}$. If an initial 
condition $\vec \xi(N)$ lies on that cone, that is if 
${h\left[ \vec \xi(N) - \vec \xi^- \right] = 0}$, then 
according to Eq.~(\ref{cone}) with ${\vec \xi_A (n) = 
\vec \xi (n)}$ and ${\vec \xi_B (n) = \vec \xi^- }$, all 
terms of the sequence $\vec \xi(n)$, generated by the 
simplified RGT of Eq.~(\ref{largen}), lie on the same cone, 
that is $h\left[ \vec \xi(n) - \vec \xi^- \right] = 0$ for 
${n=N-2, N-4, N-6}$ and so on. Hence, $\vec \xi(n)$ 
cannot reach $\vec \xi^+$ no matter how close to 1 its 
projection $f(k)$ gets. For the attractive point $\vec \xi^+$ 
does not lie on the same cone. An example of the RG 
flow that keeps cyclically coming close to $\vec \xi^+$ 
but always misses it, is shown by the blue points in 
Fig.~\ref{fig3}.

Furthermore, since the attractive fixed point lies inside the cone, for
$h\left[ \vec \xi^+ - \vec \xi^- \right] > 0$, the sequences that
start from an initial condition $\vec \xi(N)$ outside the 
cone and eventually converge on $\vec \xi^+$, so that initially 
$h\left[ \vec \xi(N) - \vec \xi^- \right] < 0$ and eventually 
$h\left[ \vec \xi(n) - \vec \xi^- \right] > 0$, must make a similar
jump to the one made by the sequence represented by the red 
points in Fig.~\ref{fig3}. They have to make the jump because 
the sequence $h\left[ \vec \xi(n) - \vec \xi^- \right]$ must change 
its sign from negative to positive for some value of the floating 
cutoff $n_0$,
\begin{subequations}
\begin{align}
&h\left[ \vec \xi (n_0) - \vec \xi^- \right] < 0 \ , \\ 
&h\left[ \vec \xi (n_0-2) - \vec \xi^- \right] > 0 \ .
\end{align}
\end{subequations}
According to Eq.~(\ref{cone}), this can only happen 
if ${\xi_1(n_0) + 1/(6gN) < 0}$.

Analogous condition to Eq.~(\ref{cone}) holds for the exact RGT, 
\begin{align}
\label{cone1}
&h\left[ \vec \xi_A (n-2) - \vec \xi_B (n-2), n-2 \right] = \nonumber\\ &= \frac{\phi_2(n) \, h\left[ \vec \xi_A (n) - \vec \xi_B (n), n \right]}{\left( \xi_{A 1}(n) + \frac{n-E}{g \phi(n)} \right)\left( \xi_{B 1}(n) +  \frac{n-E}{g \phi(n)} \right)} \ ,
\end{align}
where $h\left[ \vec v, n \right] = v_1 v_2 - \phi_1(n) v_3^2$,
which explains the exceptions mentioned in Sec.~\ref{ffp+};
that there exist sequences that do not converge on $\vec \xi^+(n)$
even if they come close.

Projection $f(k)$ can also be defined by replacing $\vec \xi^+$ and $\vec \xi^-$ in Eq.~(\ref{fk}) with the floating fixed points $\vec \xi^+(N-2k)$ and $\vec \xi^-(N-2k)$. One can derive the approximate values of $\vec \xi^+(N-2k)$ and $\vec \xi^-(N-2k)$ numerically as described before and then generate plots of $f(k)$. Such plots are similar to those presented in Fig.~\ref{fig3}. In particular, an oscillating sequence $f(k)$ analogous to the one represented by the blue points in Fig.~\ref{fig3}, is generated for an initial condition $\vec \xi(N)$ satisfying ${h\left[ \vec \xi (N) - \vec \xi^- (N), N \right] = 0}$, while a sequence with a big jump analogous to the one represented by the red points can be obtained for an initial condition  satisfying ${h\left[ \vec \xi (N) - \vec \xi^- (N), N \right] < 0}$.
%%%%%%%%%%%%%%%%%%%%%%%%%%%%%
\section{ Extension of the RG analysis }
\label{extensions}
%%%%%%%%%%%%%%%%%%%%%%%%%%%%%
Our discussion of the quartic Hamiltonian RG behavior can be 
naturally extended in different ways described in this section. 
%%%%%%%%%%%%%%%%%%%%%%%%%%%%%
\subsection{ Polynomial interactions}
\label{quadratic}
%%%%%%%%%%%%%%%%%%%%%%%%%%%%%
The first type of extension concerns Hamiltonians of the form
\begin{equation}
\label{Hgeneral}
H = -\frac{d^2}{d \varphi^2} + \sum_{i=2}^M A_i \, \varphi^i \ ,
\end{equation}
where ${M > 4}$ is a finite even number. $A_M$ is assumed positive 
for the energy spectrum to be bounded from below. $A_2$ is also 
assumed positive for the eigenstates of quadratic-oscillator 
Hamiltonian ${H_0 =  -d^2 /d \varphi^2 + A_2 \, \varphi^2}$ to 
provide the basis in which one computes the Hamiltonian matrix 
$H^\infty_{k,l}$. 

The RGT of the resulting matrix $H^\infty_{k,l}$ depends on the form 
of polynomial with ${i > 2}$  in Eq.~(\ref{Hgeneral}), which we call 
the interaction. If the interaction is even, that is if ${A_i = 0}$ for 
odd $i$, then $H$ doesn't mix even and odd eigenstates of $H_0$ 
and the number of bands of $H^\infty_{k,l}$ equals ${M+1}$. This
 implies that the cutoff flow of the Hamiltonian can be parameterized 
by a  vector $\vec \xi(n)$ of dimension ${M(M/2+1)/4}$, defined
analogously to the 3-dimensional case in Eqs.~(\ref{xidef}). 
If the potential is not even, then the number of bands equals ${2M+1}$ 
and the RGT acts on a ${M(M+1)/2}$-dimensional vector $\vec \xi(n)$. 
The RGT equation has a form ${\vec \xi(n-2) = \vec F[\vec \xi(n)]}$ in the 
case of even interactions and ${\vec \xi(n-1) = \vec F[\vec \xi(n)]}$ for 
the interactions that are not even. In the latter case the Gaussian elimination 
does not integrate out the even rows and columns of effective Hamiltonian 
independently from the odd ones. Vector $\vec F[\vec \xi]$ is a 
rational function of components of $\vec \xi$ and depends on $M$, 
${A_2, \dots, A_M}$, floating cutoff $n$ and energy $E$. One can 
study properties of the RGTs for band-diagonal Hamiltonians of 
Eq.~(\ref{Hgeneral}) following the steps analogous to the case of 
${M=4}$ discussed in the previous sections. 

We illustrate complexity of the resulting RGTs using the case of ${M = 6}$,
\begin{equation}
\label{Hsextic}
H = a^\dagger a + g \left( a + a^\dagger \right)^6 \ .
\end{equation}
The diagonal matrix elements of this Hamiltonian in the 
oscillator basis are
\begin{subequations}
\label{xisexticdef}
\begin{align}
\label{xisexticdef1}
&H_{n ;  n,n}^N = n + \xi_1(n) \left( H^N_{n,n} - n \right) ,
\\
\label{xisexticdef2}
&H_{n ;  n-2,n-2}^N = n-2 + \xi_2(n) \left[ H^N_{n-2,n-2} - \left( n-2 \right) \right] ,
\\
\label{xisexticdef3}
&H_{n ;  n-4,n-4}^N = n-4 + \xi_3(n) \left[ H^N_{n-4,n-4} - \left( n-4 \right) \right] ,
\end{align}
\end{subequations}
and the off-diagonal ones are 
\begin{subequations}
\label{xisexticdefoff}
\begin{align}
\label{xisexticdef4}
&H_{n ;  n,n-2}^N = H_{n ;  n-2,n}^N = \xi_4(n) H^N_{n,n-2} ,
\\
\label{xisexticdef5}
&H_{n ;  n,n-4}^N = H_{n ;  n-4,n}^N = \xi_5(n) H^N_{n,n-4} ,
\\
\label{xisexticdef6}
&H_{n ;  n-2,n-4}^N = H_{n ;  n-4,n-2}^N = \xi_6(n) H^N_{n-2,n-4} .
\end{align}
\end{subequations}
The dimension of $\vec \xi(n)$ is 6, instead of 4 in Eqs.~(\ref{xidef}), 
because the matrix of Hamiltonian in Eq.~(\ref{Hsextic}) has more 
bands than the quartic-oscillator matrix. The RGT expressed in terms 
of $\vec \xi(n)$ is quite complex and we do not produce it here. In 
the limit of large floating cutoffs $n$ and the number of RGT steps 
${(N-n)/2}$ much smaller than $N$, with all terms of order 
${O\left( 1/N^3 \right)}$ dropped and using
\beq
\label{d6}
d_6(n,N) \es \xi_1(n)+1/(20gN^2) \ , 
\eeq
the RGT reads
\begin{subequations}
\label{xiSSBlargen}
\begin{align}
\label{xiSSBlargen1}
&\xi_1(n-2) = \xi_2(n) - \left( \frac{9}{16} + \frac{63}{64N^2} \right) \frac{\xi_4(n)^2}
{d_6(n,N)} \ ,
\\
\label{xiSSBlargen2}
&\xi_2(n-2) = \xi_3(n) - \left( \frac{9}{100} + \frac{63}{100N^2} \right) \frac{\xi_5(n)^2}
{d_6(n,N)} \  , 
\\
\label{xiSSBlargen3}
&\xi_3(n-2) = 1 - \left( \frac{1}{400} + \frac{63}{1600N^2} \right) \frac{1}
{d_6(n,N)} \  , 
\\
\label{xiSSBlargen4}
&\xi_4(n-2) = \xi_6(n) - \left( \frac{3}{10} + \frac{21}{20N^2} \right) \frac{\xi_4(n) \xi_5(n)}
{d_6(n,N)} \  ,
\\
\label{xiSSBlargen5}
&\xi_5(n-2) = 1 - \left( \frac{1}{8} + \frac{21}{32N^2} \right) \frac{\xi_4(n)}
{d_6(n,N)} \  ,
\\
\label{xiSSBlargen6}
&\xi_6(n-2) = 1 - \left( \frac{1}{50} + \frac{21}{100N^2} \right) \frac{\xi_5(n)}
{d_6(n,N)} \  .
\end{align}
\end{subequations}
The number of floating fixed-points of this RGT and their repulsive or attractive nature requires 
studies beyond what authors have done so far. In the limit ${n \sim N \to \infty}$ one can proceed
in a simplified way, analogous to Secs.~\ref{fp+} and \ref{nearfixed}. The fixed points ${\vec \xi(n) = \vec \xi^*}$ of the RGTs in that limit can be derived numerically. For all values of $N$ and $g$ that we checked, ranging from ${N=1000}$ to ${N=10000}$ and from ${g=0.1}$ to ${g=10}$, the RGT possess 4 fixed points. 
One fixed point is attractive, one is repulsive and two other ones are mixed. In the latter, the derivative matrix $\vec F' [\vec \xi^*]$ has some eigenvalues with modulus bigger than 1 and some with modulus smaller than 1, corresponding to Wegner's relevant and irrelevant interaction terms~\cite{FJW}.

%%%%%%%%%%%%%%%%%%%%%%%%%%%%%%%
\subsection{Quartic Hamiltonians with negative\\ coefficient $A_2$}
\label{appendixB}
%%%%%%%%%%%%%%%%%%%%%%%%%%%%%%%
The second case of interest  concerns even Hamiltonians of the form 
displayed in Eq.~(\ref{Hgeneral}) with ${M=4}$ and a negative coefficient 
$A_2$. It is naturally of interest because of the spontaneous breaking 
of $Z_2$ symmetry. Without losing generality, the Hamiltonian can 
be written in the form
\begin{equation}
\label{HSSB}
H = a^\dagger a + g \left( a + a^\dagger \right)^3 + g^2 \left( a + a^\dagger \right)^4 \ .
\end{equation}
This form is derived by writing the quartic-oscillator Hamiltonian with 
negative quadratic coefficient,
\begin{equation}
\label{HSSB1}
H = -\frac{d^2}{d \varphi^2} - A \, \varphi^2 + B \, \varphi^4 \ ,
\end{equation}
in terms of variable ${\phi = \varphi - \langle \varphi \rangle}$, which describes the 
deviation of $\varphi$ from one of its so-called vacuum expectation values 
${\langle \varphi \rangle = \pm \sqrt{A/(2B)}}$. One rescales ${H \rightarrow 
H/(2\sqrt{2A})}$, introduces creation and annihilation operators via 
${\phi = (8A)^{-1/4} \left( a + a^\dagger \right)}$,  ${d / d \phi = (A/2)^{1/4} 
\left( a - a^\dagger \right)}$ and drops the constant term.  The result
of Eq.~(\ref{HSSB}) follows with ${g = \sqrt{B} / \left( 8A \right)^{3/4}}$.

The RGT for matrices of Hamiltonians obtained from Eq.~(\ref{HSSB}) 
is described using a 10-dimensional vector $\vec \xi(n)$. The diagonal 
matrix elements are described by $\xi_i$ with $i = 1, 2, . . . ,4$
\begin{subequations}
\label{xiSSBdef}
\begin{align}
\label{xiSSBdef1}
&H_{n ;  n,n}^N = n + \xi_1(n) \left( H^N_{n,n} - n \right) ,
\\
\label{xiSSBdef2}
&H_{n ;  n-1,n-1}^N = n-1 + \xi_2(n) \left[ H^N_{n-1,n-1} - \left( n-1 \right) \right] ,
\\
\label{xiSSBdef3}
&H_{n ;  n-2,n-2}^N = n-2 + \xi_3(n) \left[ H^N_{n-2,n-2} - \left( n-2 \right) \right] ,
\\
\label{xiSSBdef4}
&H_{n ;  n-3,n-3}^N = n-3 + \xi_4(n) \left[ H^N_{n-3,n-3} - \left( n-3 \right) \right] ,
\end{align}
\end{subequations}
and the off-diagonal terms by $\xi_i$ with $i = 5, 6, . . . ,10$.
\begin{subequations}
\label{xiSSBdef1}
\begin{align}
\label{xiSSBdef5}
&H_{n ;  n,n-1}^N = H_{n ;  n-1,n}^N = \xi_5(n) H^N_{n,n-1} ,
\\
\label{xiSSBdef6}
&H_{n ;  n,n-2}^N = H_{n ;  n-2,n}^N = \xi_6(n) H^N_{n,n-2} ,
\\
\label{xiSSBdef7}
&H_{n ;  n,n-3}^N = H_{n ;  n-3,n}^N = \xi_7(n) H^N_{n,n-3} ,
\\
\label{xiSSBdef8}
&H_{n ;  n-1,n-2}^N = H_{n ;  n-2,n-1}^N = \xi_8(n) H^N_{n-1,n-2} ,
\\
\label{xiSSBdef9}
&H_{n ;  n-1,n-3}^N = H_{n ;  n-3,n-1}^N = \xi_9(n) H^N_{n-1,n-3} ,
\\
\label{xiSSBdef10}
&H_{n ;  n-2,n-3}^N = H_{n ;  n-3,n-2}^N = \xi_{10}(n) H^N_{n-2,n-3} .
\end{align}
\end{subequations}
As before, $H^N_{k,l}$ denote matrix elements of $H$. 
For $n\sim N \to \infty$ and using 
\beq
\label{d10}
d_{10}(n,N) = \xi_1(n)+1/(6g^2N) \ ,
\eeq 
the RGT is obtained in the simplified form in which all terms of order 
${O\left( 1/N^2 \right)}$ are dropped.
\begin{subequations}
\label{xiSSBlargen}
\begin{align}
\label{xiSSBlargen1}
&\xi_1(n-1) = \xi_2(n) - \frac{1}{4g^2 N} \frac{\xi_5(n)^2}{d_{10}(n,N)} \ ,
\\
&\label{xiSSBlargen2}
\xi_2(n-1) = \xi_3(n) - \frac{4}{9} \frac{\xi_6(n)^2}{d_{10}(n,N)} \ ,
\\
&\label{xiSSBlargen3}
\xi_3(n-1) = \xi_4(n) - \frac{1}{36g^2 N} \frac{\xi_7(n)^2}{d_{10}(n,N)} \ , 
\\
&\label{xiSSBlargen4}
\xi_4(n-1) = 1 - \frac{1}{36} \frac{1} {d_{10}(n,N)} \ , 
\\
&\label{xiSSBlargen5}
\xi_5(n-1) = \xi_8(n) - \left( \frac{2}{3} - \frac{1}{3N} \right) \frac{\xi_5(n) \xi_6(n)}{d_{10}(n,N)} \ , 
\\
&\label{xiSSBlargen6}
\xi_6(n-1) = \xi_9(n) - \frac{1}{8g^2 N} \frac{\xi_5(n) \xi_7(n)}{d_{10}(n,N)} \ , 
\\
&\label{xiSSBlargen7}
\xi_7(n-1) = 1 - \left( \frac{1}{2} - \frac{1}{2N} \right) \frac{\xi_5(n)}{d_{10}(n,N)} \ , 
\\
&\label{xiSSBlargen8}
\xi_8(n-1) = \xi_{10}(n) - \left( \frac{2}{9} - \frac{1}{9N} \right) 
\frac{\xi_6(n) \xi_7(n)}{d_{10}(n,N)} \ , 
\\
&\label{xiSSBlargen9}
\xi_9(n-1) = 1 - \frac{1}{6} \frac{\xi_6(n)}{d_{10}(n,N)} \ , 
\\
&\label{xiSSBlargen10}
\xi_{10}(n-1) = 1 - \left( \frac{1}{18} - \frac{1}{18N} \right) \frac{\xi_7(n)}{d_{10}(n,N)} \ .
\end{align}
\end{subequations}
The evolution of the 10-dimensional vector $\vec \xi(n)$ exhibits 
similar features to the ones described at the end of Sec.~\ref{quadratic}. 
However, the case of 10 dimensions is much richer in structure 
and more difficult to fully analyze. The authors have not completed 
the full analysis.
%%%%%%%%%%%%%%%%%%%%%%%%%%%%%%%
\subsection{ Extension to quantum field theory }
\label{QFT}
%%%%%%%%%%%%%%%%%%%%%%%%%%%%%%%
The third way of extending our RG analysis of just one quartic oscillator 
concerns two or more oscillators that interact with each other. In particular, 
denoting one such oscillator variable by $\varphi_1$ and another one by
$\varphi_2$, one can consider interactions of the form $(\varphi_1 - \varphi_2)^2/(2 a^2)$.
The parameter $a$ corresponds to a discrete approximation for the spatial 
derivative $d \varphi/dx$ with $dx$ replaced by $a$. Including more oscillators
and labeling them by vectors $\vec m = (m_1,m_2, m_3)$ with integer components,
one can think about the oscillator variable $\varphi_{\vec m}$ as a quantum
field at a space grid point $\vec x = a \vec m$. The new element of such setup 
for studying scalar theory on the spacial lattice is that the oscillators at each 
and every point $\vec x$ are cut off in ultraviolet by a cutoff $N$ and involve 
a corresponding vector $\vec \xi (\vec x, N)$. 

Consider the Hamiltonian for two oscillators $\varphi_1$ and $\varphi_2$ 
coupled by the term 
\begin{equation}
\label{phiphi}
(\varphi_1 - \varphi_2)^2/(2 a^2) 
=
\varphi_1^2 /(2 a^2) + \varphi_2^2/(2 a^2) 
-
\varphi_1 \varphi_2/ a^2 \ .
\end{equation}
The first two terms on the right-hand side individually modify the quadratic 
terms of each of the oscillators. The only term that couples the oscillators 
is the third one. It is capable of exciting or de-exciting both oscillators 
by 1, or exciting one oscillator by 1 and de-exciting the other one by 1. 
The interaction contributes to the matrix elements of the Hamiltonian, 
denoted in a self-explanatory way by $H^N_{k_1,k_2; l_1,l_2}$.
The matrix elements involve two 3-dimensional vectors $\vec \xi(1,N)$ 
and $\vec \xi(2,N)$. However, the number of eigenstates of definite free oscillator 
energy $K$ in units of the modified $\hbar \omega$ is $K+1$. The 
Gaussian elimination becomes complicated. The authors have not identified 
any simple way to carry it out. The RG analysis of a similar Hamiltonian 
matrix for an entire grid of oscillators not only involves the vector $\vec \xi(N)$ 
as many times as there are grid points but also must deal with a significant 
increase in degeneracy. Implications of setting up a regulated quantum field 
theory this way, with finite parameters $N$ and $a$, require investigation, 
including the issue of how to approach the limits of $N \to \infty$ and 
${a \to 0}$ simultaneously. 
%%%%%%%%%%%%%%%
\section{Conclusion and outlook}
%%%%%%%%%%%%%%%
This article establishes that the basic quantum-physics system of 
an oscillator with a quartic interaction term, exhibits the spiral cutoff 
dependence in the Wilsonian renormalization group procedure. Its 
Hamiltonian matrix in the basis of harmonic oscillator eigenstates 
is band-diagonal and its cutoff flow involves only the several matrix 
elements between the basis states of highest allowed oscillator energy. 
These evolving matrix elements are parameterized in terms of components 
of the 3-dimensional vector that spirals towards a slowly drifting, 
attractive fixed point as the cutoff decreases, with some exceptions. 

Renormalization of band-diagonal Hamiltonian matrices was discussed 
in the past, see {\it e.g.}~\cite{[See App. B in ]GW93}. However, to 
the best of authors' knowledge the possibility of a cutoff spiral flow 
in them was not noted to exist. Such possibility is also difficult to 
note in case of the quartic oscillator, because the initial spiral contraction 
rate can be so large that it causes an illusion of a rapid convergence to 
an apparent fixed point. 

Further renormalization-group studies of polynomial interactions are 
worth pursuing for the purpose of understanding their cutoff dependence,
see Sec.~\ref{extensions}. In particular, studies of polynomial interactions 
of order 4 with negative quadratic term are desired for understanding the 
dynamics of spontaneous breaking of $Z_2$ symmetry. Generalization of 
the RG studies to coupled quartic oscillators provides a way to investigate 
the role of cutoffs in Hamiltonian formulation of quantum field theory. 

It should be stressed that this paper does not discuss corrections 
due to the energy eigenvalues $E$ that enter the Gaussian elimination. 
An example that illustrates a method for analyzing such dependence 
using expansion in powers of the ratio $E/n$, where $n$ stands for
the floating cutoff, is available in~\cite{SDGKGW}.

\vskip.2in
{\bf Acknowledgement}

The authors thank Kamil Serafin for his suggestion that 
one could consider a slowly varying point, instead of a 
fixed one, to describe results obtained in Ref.~\cite{KWojcik}.

\nocite{*}

\bibliography{OscillatorGirgusGlazek20240426}% Produces the bibliography via BibTeX.
\end{document}